# Nanofabrication of high *Q*, transferable, diamond resonators

*Blake Regan[1], Aleksandra Trycz[1], Johannes E. Fröch[1], Otto Cranwell Schaeper[1], Sejeong Kim[1], Igor Aharonovich[1,2]*

[1] School of Mathematical and Physical Sciences, University of Technology Sydney, Ultimo, New South Wales 2007, Australia
[2] Centre of Excellence for Transformative Meta-Optical Systems, University of Technology Sydney, Ultimo, NSW 2007, Australia

*igor.aharonovich@uts.edu.au

**Abstract**

*Advancement of diamond based photonic circuitry requires robust fabrication protocols of key components – including diamond resonators and cavities. Here, we present 1D (nanobeam) photonic crystal cavities generated from single crystal diamond membranes utilising a metallic tungsten layer as a restraining, conductive and removable hard mask. The use of tungsten instead of a more conventional silicon oxide layer enables good repeatability and reliability of the fabrication procedures. The process yields high quality diamond cavities with quality factors (Q-factors) approaching $1\times10^4$. Finally, we show that the cavities can be picked up and transferred onto a trenched substrate to realise fully suspended diamond cavities. Our fabrication process demonstrates the capability of diamond membranes as modular components for broader diamond based quantum photonic circuitry.*

The advent of commercially available, high-quality single crystal diamonds boosted a decade of research rendering diamond as one of the most studied platforms for novel quantum technologies[1-5]. The availability of diamond colour centers (e.g. nitrogen – vacancy centres[6] or the more recent group IV defects[7]) that can be utilised as solid state qubits motivated the advance of diamond nanofabrication processes. Of particular interest is engineering of thin diamond membranes[8-13] and consequently photonic resonators including photonic crystal cavities (PCCs), microrings and waveguides[14-22]. These devices are needed for both integrated diamond photonics on a single chip, and an improved collection efficiency of photons emitted from the embedded emitters. Coupling the emitters to PCCs also constitutes the realisation of a Purcell enhancement that is needed to enhance the emission rate of photons into the zero photon line (ZPL), thus achieving a coherent photon emission[23, 24].

Several approaches toward generating optical resonators in diamond have already been established. Vertical isolation in bulk diamond has been achieved with angled reactive ion etching (RIE) employing a Faraday cage[25] or angled ion beam etching[26]. Nanobeam cavities were among the structures generated by this technique with quality factors approaching 300,000 at the infrared spectral range[25]. However, the nature of the angled etching technique inherently imposes a triangular cross section to the fabricated structures. The use of a quasi-isotropic undercut has also been demonstrated, realizing a wide variety of vertically isolated photonic structures[19, 27, 28].

For many applications, including integration with fibre cavities and on chip devices, planar geometries are required to satisfy the requirement of optical isolation of the diamond device through refractive index contrast while minimising the size of the diamond components, which is challenging with bulk diamond fabrication methods. To circumvent these issues, diamond membranes can be adopted for nanofabrication[10, 29-31], which are generated either by etching down a thick (~ 30 μm) laser cut diamond slab or by employing a lift off process from a commercially available diamond material. The latter method is advantageous as it reduces the wastage that is otherwise present in bulk diamond manufacturing. The ion implantation process causes damage to the diamond, which can reduce the effectiveness of photonic devices, however this can be avoided through an overgrowth step which generates pristine diamond[32].

Several nanofabrication approaches to create photonic cavities from diamond membranes have been established which cater to challenges in thin film fabrication. Specifically, a method using a hydrogen silsesquioxane (HSQ) as a hard mask has been broadly adopted and produced a wide variety of optical resonators[11, 33-35], including nanobeam cavities with quality factors up to ~7,000 within the visible range[36]. Nevertheless, HSQ requires a far higher dose (1,500 µC/cm$^2$) than typical electron beam (E-beam) resists, which can lead to charging artefacts caused by either a larger spot size from a higher beam current, or deflection from charge build-up while lingering on an insulating material. One of the solutions is to employ transferable silicon-based hard masks[37], which utilise the more mature fabrication protocols in silicon nanofabrication and minimise the detrimental effects of charging and resist thickness variations in lithographic processes. However, the utility of this process is limited by the necessity of dangerous chemicals, including hydrofluoric acid, and further the low reliability of pick and place masking methods with unrestrained diamond membranes, which often remove the membrane along with the mask due to adhesion. Note that while the use of HF is considered a commonplace, it is still a highly hazardous chemical, and is the most common and reliable method of removing the silica hard mask generated by HSQ. By omitting the use of HF, we reduce the dependence on hazardous processes while producing similar results, improving the accessibility of nanofabrication techniques.

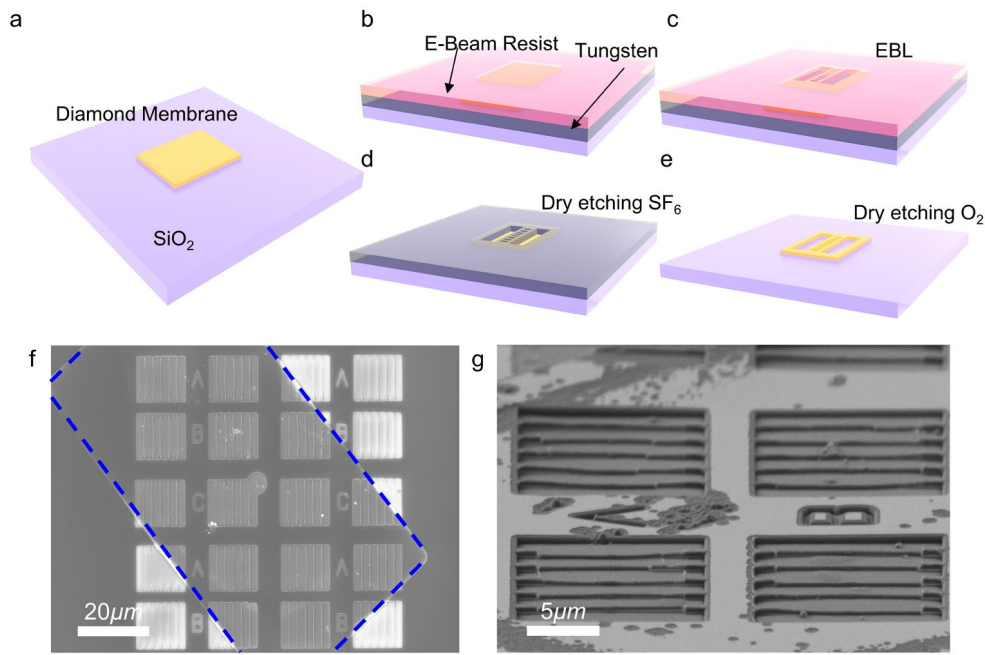

*Figure 1. Concept of tungsten hard masking fabrication process. a) A single crystal, 300 nm thick diamond membrane, placed on SiO$_2$ substrate. b) Deposition of 30 nm of tungsten by magnetron sputtering system, followed by spin coating and curing of CSAR EBL resist. c) Exposure and development of the pattern. d) Photoresist pattern transferred to the tungsten hard mask through SF$_6$ dry etching. e) Dry O$_2$ etching to transfer the pattern to the diamond membrane. f) SEM image of the fabricated photonic cavities. The blue outline indicates the diamond membrane with scale bar corresponding to 20 µm. g) SEM image of the arrays of the diamond nanobeam photonic crystal cavities. Scale bar is 5 µm.*

In this work, we demonstrate an alternative diamond nanofabrication protocol utilising a tungsten hard mask layer, to fabricate photonic resonators from diamond membranes. Our work matches the urgent need for robust and cost-effective methodologies to fabricate high quality, nanoscale photonic devices from single crystal diamond membranes. The tungsten metallic hard mask serves three purposes: first, it acts as a uniform conductive layer during electron beam lithography (EBL) process. Second, it acts as a restraining layer to minimize the probability that the membrane will detach from the substrate. Finally, the tungsten also acts as a hard mask through the highly selective etch chemistries to produce the final diamond resonators. The alternative fabrication approach utilises the physical, chemical and conductive properties of the masking material to enable a reliable fabrication processes while retaining a quality factor comparable to other methods.

Figure 1 shows the procedure of the fabrication process. First, diamond membranes containing colour centres (SiV in this case) were fabricated using a method described previously[11], and are transferred onto a clean silicon oxide substrate (Figure 1a). The sample then undergoes deposition of a 30 nm tungsten layer in a magnetron sputter deposition system at 100W under 50 sccm argon plasma, which generates a homogeneous 30nm metal layer across the sample. Following the tungsten layer deposition, electron beam lithography resist (E-Beam Resist AR-P 6200 CSAR-Allresist) is spin coated on the substrate at 3,000 rpm, forming an approximately 500 nm polymer resist layer for EBL as shown in Figure 1b. The tungsten layer acts as a uniform conductive layer, reducing the charging effect during EBL on diamond membranes, which allows patterning of small photonic crystal features below 50nm. Arrays of crystal cavities are patterned over the diamond membrane, as demonstrated in Figure 1c. The CSAR electron beam resist is developed at 5°C for 13 seconds in a development solution, where the membrane is secured on the substrate by both the resist and metallic tungsten layers.

The transfer of the pattern onto the diamond is done *via* a two-step dry etching process. The first etch is a dry $SF_6$ etch which transfers the resist pattern into the tungsten. This etch is conducted at 5mTorr under 10sccm of $SF_6$ and 4sccm of Ar, with 50w ICP power and 75W RIE for 45 seconds. Given the resistivity of CSAR and the rapid etch rate of tungsten under fluorine, the pattern is transferred to the hard mask with minimal risk of complete removal of the resist, as schematically illustrated in figure 1d. Figure 1e represents the $O_2$ inductively coupled plasma reactive ion etching (ICP-RIE) transfer of the pattern from tungsten into diamond, this process also includes the removal of the E-beam resist. Note that the tungsten layer acts as the hard mask during oxygen plasma and forms a tungsten oxide film. The etch parameters were 45sccm of $O_2$ at 10mTorr with 500W ICP and 100W RIE for 5 minutes. The durability of tungsten oxide under oxygen etch conditions enables a high selectivity for diamond etch (~ 30 nm of sputtered material to completely etch through the 300 nm diamond membrane). The tungsten oxide film and remaining metallic tungsten were removed with ~500 μL droplet of hydrogen peroxide. Scanning electron microscope (SEM) images of the final structures demonstrate the outcomes of this fabrication procedure (Figure 1f, 1g). After mask removal, the samples were transferred freely to a trenched $SiO_2$ substrate, which had been patterned through standard photolithography, employing an $SF_6$ etch to transfer the pattern into the silicon. During the transfer process, the membrane is released either by direct contact to the substrate or by placing a droplet of water on the substrate surface and dipping the probe tip in, releasing the membrane. Additional SEM images are shown in the supporting information.

To study the optical properties of the diamond cavities, photoluminescence (PL) measurements were conducted under a continuous-wave 532 nm laser excitation at room temperature using a confocal microscope. Figure 2a and 2b show SEM images with tilted view of the fabricated photonic devices. 1D photonic crystal cavities were modelled and simulated *via* finite-difference time-domain (FDTD) method using Lumerical software, generating the mode profile as shown in Figure 2c. A ladder-like design with rectangular air holes is adopted to maximize the effective refractive index contrast. The calculated Q-factor from simulation is determined to be ~ $2\times10^5$ and the mode volume corresponds to ~ $2.8(\lambda/n)^3$

Figure 2d depicts the finding as a histogram of *Q*-factors. Cavities with low quality factors were found to correlate to cavities with damaged side walls, those with incomplete or warped structure due to a proximity to the membrane edge. Figure 2e shows a spectrum of a 1D nanobeam cavity. The characteristic SiV ZPL at ~738 nm is visible, along with a cavity mode augmenting the SiV peak at 735 nm, with $Q$ of ~1,600 and a second mode at ~700nm with a Q~1750. Figure 2f demonstrates cavity designs that targeted alternative colour centers. The mode at 621 nm correlates to the known emission centers of SnV with ZPL at ~ 620 nm. The cavities with the highest quality factors were measured at the near infrared range, with a cavity resonance at ~ 775 nm, exhibiting $Q$ ~ 8,400 (figure 2g).

Overall, more than ~75% of the fabricated cavities exhibited cavity modes with Q-factors over ~ 500, emphasizing the promising nanofabrication protocol with a tungsten mask. The distribution of Q factors is a combination of several factors. Damage to the structure sidewalls comes from imperfections in transferring caused by micro masking sputtering in the initial etching stages. This issue can be addressed in iterative development of this technique, focusing on improving two key elements. First is the development of the initial EBL pattern, which can be subject to footing and development artefacts which cause difficulty in the mask transfer process. The second issue is the generation and distribution of tungsten oxide, which can cause roughening of etched side walls. This issue can be addressed with a controlled temperature stage and chemical composition in the RIE etching stage. Additional, though minor contribution to the range of quality factors comes from variations in the membrane before the cavity fabrication, this may include variations in thickness, emission, surface condition and surface contamination. Each of these factors can cause variance in the quality of the outcome.

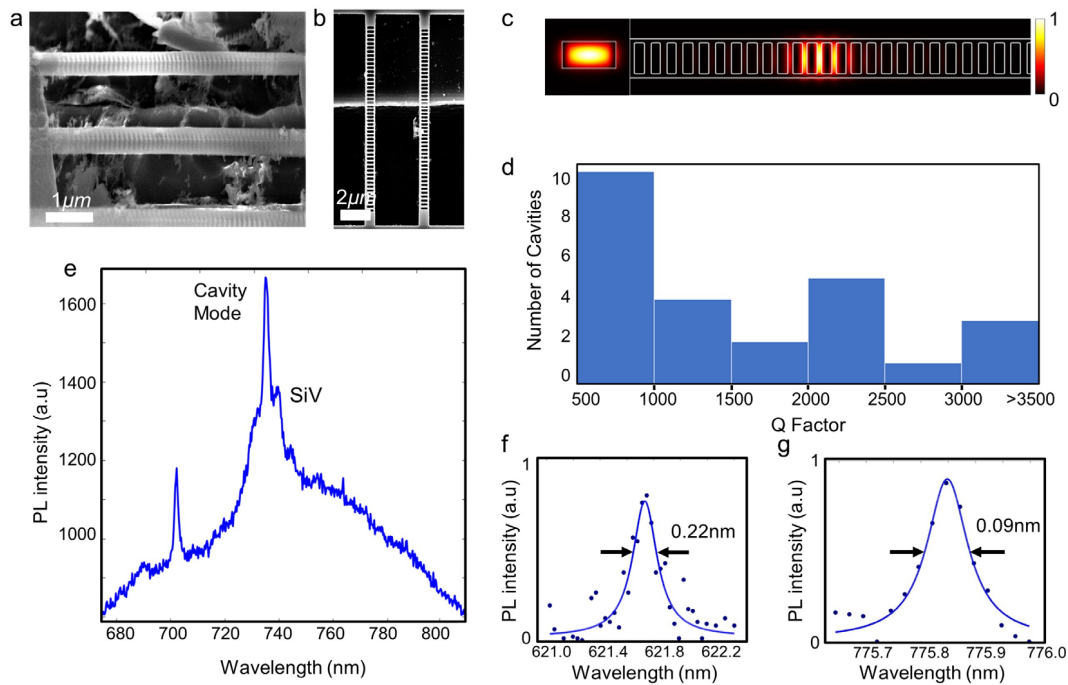

*Figure 2. Characterisation of diamond nanobeam cavities. (a, b) side and top view SEM images of the fabricated diamond nanobeam photonic crystal cavities on a native oxide silicon substrate with. Scale bars correlate to 1 µm and 2 µm, respectively. c) Electric field intensity profile of the fundamental nanobeam photonic crystal cavity modes. d) Histogram representing the distribution of Q-factors from the 1D nanobeam cavities. e) PL spectrum from nanobeam cavity depicting an optical cavity mode in proximity to characteristic SiV colour center emission peak at ~ 738 nm. f, g) High resolution spectrum of a cavity mode at ~ 620 nm and ~ 775 nm, corresponding to Q-factor of ~2,800, and ~8,400, respectively.*

To further demonstrate the capabilities of the tungsten fabrication technique, nanobeam PCCs were fabricated and subsequently transferred post fabrication onto a trenched substrate. Use of a trenched substrate is ideal to achieve optical isolation of the cavities that should result in an improved optical confinement. The transfer process depicted in Figure 3a utilises a probe tip to capture and lift the diamond membrane, allowing it to be carried between samples. They key characteristic of the probe is a hard material with a small diameter tip (~ to the approximate thickness of the diamond membrane). The diamond cavities are sufficiently robust, allowing the probe tip to be manually handled. An optical microscope with a long objective length was utilised to improve visibility. The membrane is adhered to the probe with electrostatic forces and can be released on the target substrate with a direct contact. Alternatively, the membrane may be released from the tip by adding a water droplet to the target substrate, and simply contacting the membrane to the droplet, releasing it safely to the substrate.

Figure 3b and 3c show nanobeam photonic cavities before and after the transfer, respectively. The transfer process caused no degradation to the diamond cavities. The SEM images demonstrate no observable damage to the diamond nanobeam cavities. Some residual diamond material present in figure 3b was removed during the transfer process, overall improving the quality of the diamond structure. Note, that the transfer process under ambient condition is robust, and can be easily automated. Importantly, it does not require sophisticated piezo elements or high vacuum. The range of target substrates is also not limited and those can also be structured dielectric substrates, waveguides or other cavities.

A comparison from sets of identically fabricated diamond cavities (i.e. same cavities suspended and non suspended) is shown in figure 3d. The results clearly indicate an increase in quality factor due to the improved optical confinement in the suspended cavities. The markers represent data points of th seven nanocavities before and after transfer, with the box and whisker plot indicating the range of quality factors in each condition.

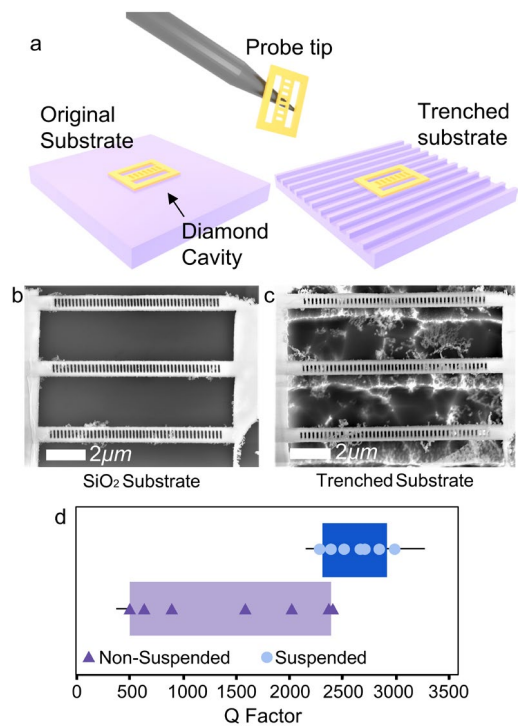

*Figure 3. Transfer of a fabricated diamond resonator. a) Schematic Illustration, the diamond cavities are lifted-off using a handheld probe tip, then released onto a substrate of choice –a trenched substrate in this case. b) Fabricated diamond membrane on original substrate after tungsten removal. c) membrane transferred and flipped onto a trenched Silicon substrate. Scale bars both represent 2μm. d)*

*Distribution plot of Q-factors comparing the same seven suspended and non-suspended photonic cavities. Open circles/triangles indicate the Q of the individual cavities.*

Finally, we qualitatively compare the tungsten fabrication method with other techniques used to fabricate diamond devices. To provide a meaningful comparison, we focus only on 1D nanobeam cavities at the visible - near infrared spectral range as those are the most meaningful for interface with diamond color centres. The results are shown in Table 1. Overall, our method is on par with other existing methodologies in terms of cavity *Qs* achieved. As expected, quasi isotropic etching of bulk diamond devices provides improved results, but the method is challenging to employ for planar architectures. A specific approach that worth a more detailed comparison is the one that employs HSQ as the hard mask and diamond membranes, as it is often considered a "single step" fabrication protocol. This masking method has demonstrated *Q* factors of up to 7,000 within the optical range. PMMA or a similar adhesive can be used to glue the membrane to the surface, limiting the possibility of the membrane lifting off during the wet chemical development process. However, the high dosage requirement of HSQ exacerbates the effect of charge artefacts which alter the mask structure, making the repeated use of HSQ unreliable and highly susceptible to any non-ideal exposure conditions.

The method presented here addresses the drawbacks of diamond membrane fabrication. The utilisation of a tungsten hard mask simultaneously allows wet chemical processes by acting as a restraining layer, but in addition, enables a uniform conductive layer to improve the viability of EBL.

| Reference | Fabrication method | Substrate | Quality factor | Resonant wavelength (nm) |
|---|---|---|---|---|
| [25] | Faraday cage angle etch | Bulk single crystal diamond | 8300 | 734 |
| [19, 22] | Quasi isotropic undercutting | Bulk single crystal diamond | 14700 | 637 |
| [38] | Hybrid RIE FIB undercut | Bulk Single crystal diamond | 1710 | 568 |
| [36] | HSQ/SiO2 hard mask on diamond membrane | Single crystal diamond membrane suspended on PMMA | 7000 | 636 |
| [37] | Transferrable Silicon based hard mask | Single crystal diamond membrane | 9900 | 776 |
| This Work | Tungsten hard mask | Single crystal diamond membrane | 8400 | 775 |
| This Work | Tungsten hard mask | Single crystal diamond membrane | 2800 | 620 |

*Table 1. Comparison of fabrication techniques found in recent literature utilised to generate nanobeam cavities from single crystal diamond.*

In conclusion, we have demonstrated a promising method to fabricate photonic crystal cavities from diamond membranes using a metallic tungsten hard mask. The primary feature which enables fabrication is the tungsten mask along with the resist layer, which restrain the membrane to the substrate while acting as a uniform conductive layer. This eliminates charging issues on the diamond and diamond-substrate interface, which is a key factor in improving nanoscale patterning resolution. While there are several other metals (e.g titanium) which could be employed in a similar manner, the use of tungsten as a hard mask is based on high selectivity with respect to the diamond.

Etching under $SF_6$ enables a rapid transfer of the pattern to the hard mask, and the high resistivity to $O_2$ based etching used to transfer the pattern into the diamond. Furthermore, the resultant tungsten oxide layer formed during the oxygen plasma etching can either be removed under a further $SF_6$ dry etch or with a droplet of hydrogen peroxide. The use of $H_2O_2$ eliminate the necessity of using dangerous chemicals such as hydrofluoric acid that is commonly used in silicon oxide based hard masks. Employing the tungsten fabrication method, high quality 1D diamond resonators were demonstrated, and the devices were easily transferred onto a substrate of choice. The presented technique is promising for large scale diamond nanofabrication and generation of planar diamond photonic devices.

**Supplementary material**
The supplementary material includes two close up images of the photonic crystal cavities.

**Acknowledgements**
The authors acknowledge the funding from the Australian Research council (via DP180100077 and CE200100010), the Asian Office of Aerospace Research and Development grant: FA9550-19-S-0003. The authors would like to thank the ANFF UTS hub for access to their facilities. The authors thank Kumar Ganesan for access to ion implantation facilities. The authors would like to thank Milad Nonahal for his fruitful discussions.

**Data availability statement**
All data is available upon request from the corresponding author